\documentclass[10pt,twocolumn,english,aps,prd,nofootinbib,preprintnumbers]{revtex4}

\usepackage[latin9]{inputenc}
\usepackage{graphicx}
\usepackage{amsmath}
\usepackage{amssymb}
\usepackage{amsfonts}
\usepackage{lscape}
\usepackage{hyperref}
\usepackage{url}
\usepackage{color}
\usepackage{bm}
\usepackage{tabularx}
\usepackage{mathtools}
\newcommand{\be}{\begin{equation}}
\newcommand{\ee}{\end{equation}}
\newcommand{\ba}{\begin{eqnarray}}
\newcommand{\ea}{\end{eqnarray}}
\newcommand{\nn}{\nonumber}

\renewcommand{\[}{\begin{equation}}
\renewcommand{\]}{\end{equation}}
\def\be{\begin{equation}}
\def\ee{\end{equation}}
\def\bea{\begin{eqnarray}}
\def\eea{\end{eqnarray}}
\def\eqi{\begin{equation}}
\def\eqf{\end{equation}}
\def\eqia{\begin{eqnarray}}
\def\eqfa{\end{eqnarray}}

\makeatother

\begin{document}

\preprint{IFT-UAM/CSIC-19-6}

\title{Can the homogeneity scale be used as a standard ruler?}

\author{Savvas Nesseris}
\email{savvas.nesseris@csic.es}

\author{Manuel Trashorras}
\email{manuel.trashorras@csic.es}

\affiliation{Instituto de F\'isica Te\'orica UAM-CSIC, Universidad Auton\'oma de Madrid,
Cantoblanco, 28049 Madrid, Spain}

\date{\today}

\begin{abstract}
The Universe on comoving scales larger than 100 Mpc/h is assumed to be statistically homogeneous, with the transition scale where we have $1\%$ deviation from homogeneity being known as the homogeneity scale $R_H$. The latter was recently proposed in Ntelis et al. (2018) [arXiv:1810.09362] to be a standard ruler. In this paper we perform a comparison between the baryon acoustic oscillations and the homogeneity scales $R_H$, assuming a spatially flat universe, linear cosmological perturbation theory and the cosmological constant cold dark matter $\Lambda$CDM model. By direct theoretical calculations, we demonstrate that the answer to the question of whether the homogeneity scale can be used as a standard ruler, is clearly negative due to the non-monotonicity of the homogeneity scale $R_H$ with the matter density parameter $\Omega_{m0}$ of the $\Lambda$CDM model. Furthermore, we also consider the effect of redshift space distortions and we find that they do not break the degeneracies but instead only change the value of $R_H$ by a few percent. Finally, we also confirm our findings with the help of an N-Body simulation.
\end{abstract}
\keywords{cosmology, dark energy, large scale structure}

\maketitle

\section{Introduction \label{sec:intro}}
One of the main pillars of modern relativistic cosmology is the cosmological principle, that is the assumption that the Universe is homogeneous and isotropic on large (cosmological) scales, usually taken be on the order of $\sim100$ Mpc/h \cite{Martinez:1998yp}. On smaller scales we observe structures that are formed as a result of gravitational collapse, which are in good agreement with the predictions of the cosmological constant and cold dark matter model known as $\Lambda$CDM \cite{Aghanim:2018eyx}. Furthermore, the $\Lambda$CDM model passes many of the standard cosmological tests and is currently favored by most of the recent cosmological observations such as Planck \cite{Aghanim:2018eyx}, but it also faces some challenges that could be resolved in the near future with data coming from current and upcoming surveys \cite{Bullock:2017xww, Perivolaropoulos:2011hp,Kazantzidis:2018rnb,Primack:2015kpa,BoylanKolchin:2011dk,Amendola:2012ys,Sapone:2014nna}.

The assumption of statistical homogeneity has already been tested in the past by several surveys like WiggleZ \cite{Scrimgeour:2012wt}, BOSS CMASS \cite{Ntelis:2016suu} where it was found that indeed the Universe at scales of roughly $\sim100$ Mpc/h it starts to become homogenous and is indeed in agreement with the cosmological principle, see also \cite{Alonso:2013boa,Alonso:2014xca} for measuring the transition to statistical homogeneity with photometric redshift surveys. Furthermore, an analysis of the SDSS galaxy sample it was also found in Ref.~\cite{Labini:2008ct} that statistical homogeneity occurs at roughly $\sim 100$ Mpc/h, but large-scale fluctuations in the luminous red galaxies can be found on scales as large as $\sim 500$ Mpc/h \cite{Wiegand:2013xfa}. Recently, the effects of inhomogeneities on the baryon acoustic oscillations (BAO) standard ruler, with offsets $\lesssim 100$ Mpc/h were also detected, something which could provide another test for the $\Lambda$CDM model \cite{Roukema:2014tta}. Also, a big structure of size $\sim 500$Mpc has been detected in the SDSS catalog \cite{Clowes:2012pn}, which the authors suggest is in contradiction to
homogeneity on lower scales, while Ref.~\cite{Nadathur:2013mva} argues that the structure is compatible with statistical homogeneity.

Usually the way the scale of statistical homogeneity is estimated is via the normalized count-in-spheres method, defined as the number $N(<r)$ of objects in a galaxy catalog inside a spherical shell of radius $r$, compared to the number of objects $N_{\textrm{random}}(<r)$ in a similar shell of a mock uniformly random catalog, or equivalently
\be
\mathcal{N}(<r)=\frac{N(<r)}{N_{\textrm{random}}(<r)}, \label{eq:defNr}
\ee
where the term in the denominator is used to account for geometric effects. Assuming a spatially flat universe, then in a properly made and homogeneous mock catalog the number of objects will scale as $N_{\textrm{random}}(<r)\propto r^3$, while in general the number of objects in the real catalog will be scale as $N(<r)\propto r^{\mathcal{D}_2}$, where $\mathcal{D}_2$ is the fractal index, which for a purely homogeneous catalog takes the value $\mathcal{D}_2=3$ or asymptotes to that value at large scales for a real catalog. Clearly, the counts in spheres do not follow these analytical relations in individual regions on those scales; they only tend to follow them when averaged over what is for simplicity assumed to be a statistical ensemble of realizations of a region on the relevant scale.

The fractal index $\mathcal{D}_2$ can be calculated via the logarithmic derivative of the count-in-spheres number, given by Eq.\eqref{eq:defNr}:
\be
\mathcal{D}_2(r)=3+\frac{d \ln \mathcal{N}(<r)}{d \ln r}. \label{eq:defD2}
\ee
The fractal index has already been constrained by the WiggleZ collaboration to be within $1\%$ of $\mathcal{D}_2=3$ at distances larger than $71\pm 8~\textrm{Mpc}/h$ \cite{Scrimgeour:2012wt}. From Eq.~\eqref{eq:defD2} we can also define the homogeneity scale as the scale $R_H$ at which the Universe deviates at $1\%$ from homogeneity, i.e.
\bea
\mathcal{D}_2(R_H)=2.97~.\label{eq:d2H}
\eea
Recently, Ref.~\cite{Ntelis:2018ctq} introduced the idea that the homogeneity scale $R_H$, defined via Eq.~\eqref{eq:d2H} could be used as a standard ruler similar to that of the BAO, something which in principle is an attractive idea given that $R_H$ can be easily determined from galaxy catalogs. In the next sections we will expand on this idea, with several analytic calculations for the case of a spatially flat $\Lambda$CDM model. We will then demonstrate that $R_H$ suffers from degeneracies due to the fact that $R_H$ is not a one-to-one function of the matter density parameter $\Omega_{m0}$, something that potentially limits its applicability in realistic analyses.

The layout of our paper is as follows: In Sec.~\ref{sec:framework} we present the theoretical framework and the numerical results of our analysis of the homogeneity scale, then in Sec.~\ref{sec:nbody} we present a comparison of our method and results from a single N-body cosmological simulation and finally in Sec.~\ref{sec:conclusions} we summarize our analysis and present our conclusions.

\section{Theoretical framework and numerical results\label{sec:framework}}
In this section we briefly describe our theoretical framework, assuming everywhere a spatially flat $\Lambda$CDM model, i.e. $\Omega_{K0}=0$. In order to estimate the homogeneity of a galaxy catalog we implement the fractal dimension $\mathcal{D}_2(r)$ given by \cite{Ntelis:2018ctq}:
\bea
\mathcal{D}_2(r)=3+\frac{d \ln}{d \ln r}\left[1+\frac{3}{r^3}\int_0^r\xi(x)x^2 dx\right],\label{eq:d2xi}
\eea
where $\xi(r)$ is the two point galaxy correlation function and this formula is only valid for a flat universe, which however is not mentioned in \cite{Ntelis:2018ctq}. Clearly, since the correlation function can also be written via the number of objects in shells of radius $r$ via the counts-in-spheres method, both Eqs.~\eqref{eq:defD2} and \eqref{eq:d2xi} are totally equivalent. Then, using linear theory and ignoring redshift-space distortions (RSDs) for now, the correlation function $\xi(r)$, in a spatially flat and simply-connected space, can be written as
\bea
\xi(r)=\frac{1}{2\pi^2}\int_0^\infty j_0(k r) k^2 P(k) dk,\label{eq:corfunc}
\eea
where $P(k)$ is the matter power spectrum and the spherical Bessel function $j_0(x)$ is given by $j_0(x)=\frac{\sin(x)}{x}$. Since we are interested only in large scales, for now we only consider the linear part of the spectrum.

Using Eq.~\eqref{eq:corfunc} in \eqref{eq:d2xi} we find that we can rewrite the fractal dimension $\mathcal{D}_2(r)$ as
\bea
\mathcal{D}_2(r)=3+\frac{d \ln}{d \ln r}\left[1+\frac{3}{r}\int_0^\infty \frac{1}{2\pi^2}~j_1(k r) k P(k) dk\right],\label{eq:d2xi1}
\eea
where the spherical Bessel function $j_1(x)$ is given by $j_1(x)=\frac{\sin(x)-x\cos(x)}{x^2}$.

With the above theoretical framework, we can now calculate the theoretical predictions of the homogeneity and BAO scales. The former can be calculated directly via Eqs.~\eqref{eq:d2H} and \eqref{eq:d2xi1}. On the other hand, we can extract the BAO scale from the galaxy correlation function directly by interpolating $\xi(r)$ and then finding the local maximum at $r>80 \textrm{Mpc}/h$. Unless specified otherwise, we have used a cosmology corresponding to the best-fit Planck 18 flat $\Lambda$CDM model given by $\Omega_{b0}=0.0482205$, $\Omega_{c0}=0.2628795$, $\Omega_{m0}=0.3111$, $\Omega_{\Lambda0}=1-\Omega_{m0}=0.6889$, $\Omega_{K0}=0$ and $h=0.6766$  \cite{Aghanim:2018eyx}. For the theoretical prediction of the matter power spectrum $P(k)$ we use the CAMB code by A.~Lewis et al. \cite{Lewis:1999bs}.

\begin{figure*}[!t]
	\centering
	\includegraphics[width=0.49\textwidth]{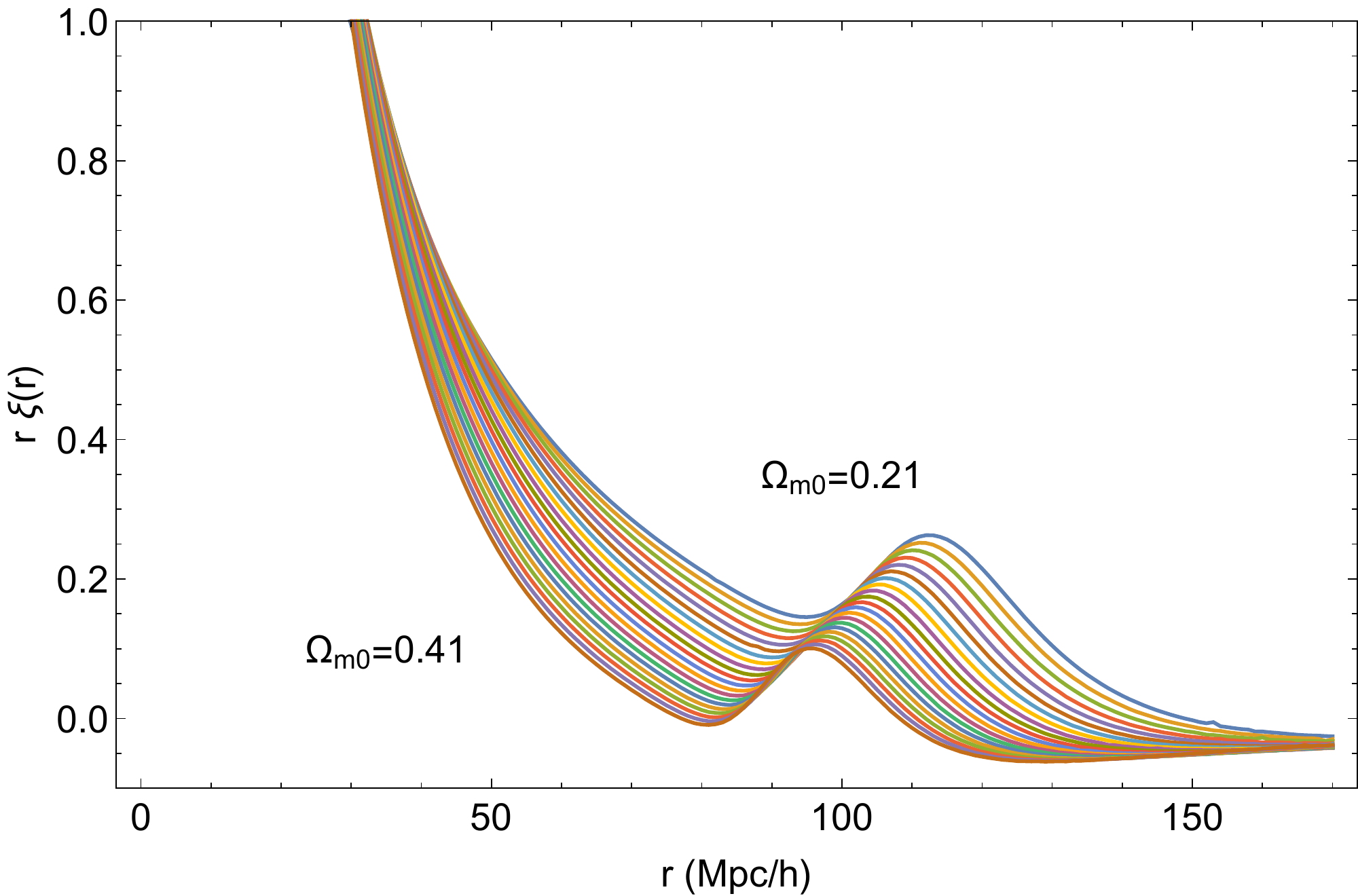}
    \includegraphics[width=0.49\textwidth]{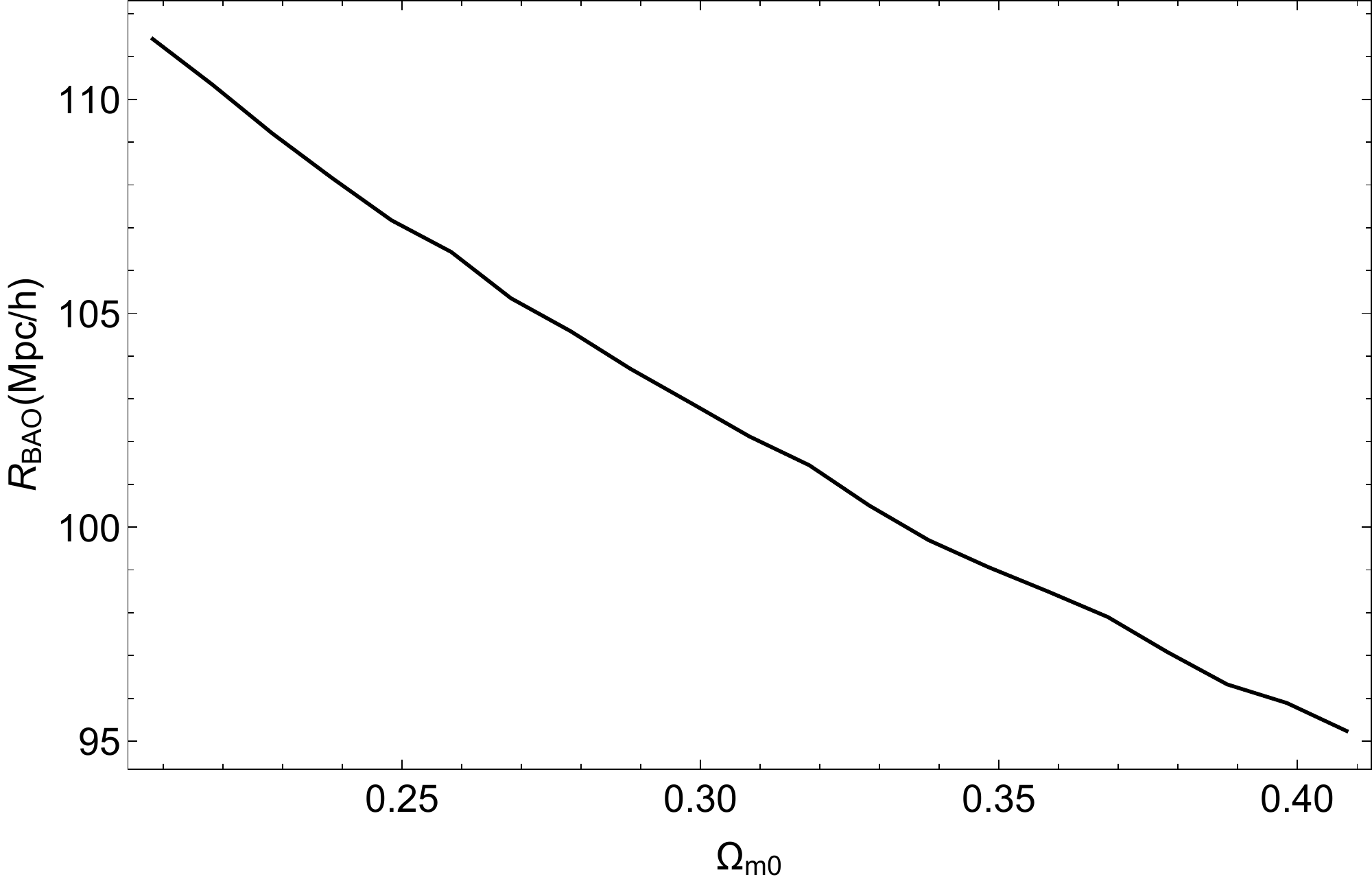}
	\caption{The galaxy correlation function $\xi(r)$ (left) for $\Omega_{m0}\in[0.21,0.41]$ and the BAO scale $R_{BAO}$ (right) as a function of the present day matter density $\Omega_{m0}$.  As can be seen, the BAO scale is a monotonically decreasing function for all values of $\Omega_{m0}$ in that range. The cosmology used is described in the text.}
	\label{fig:bao}
\end{figure*}

\begin{figure*}[!t]
	\centering
	\includegraphics[width=0.49\textwidth]{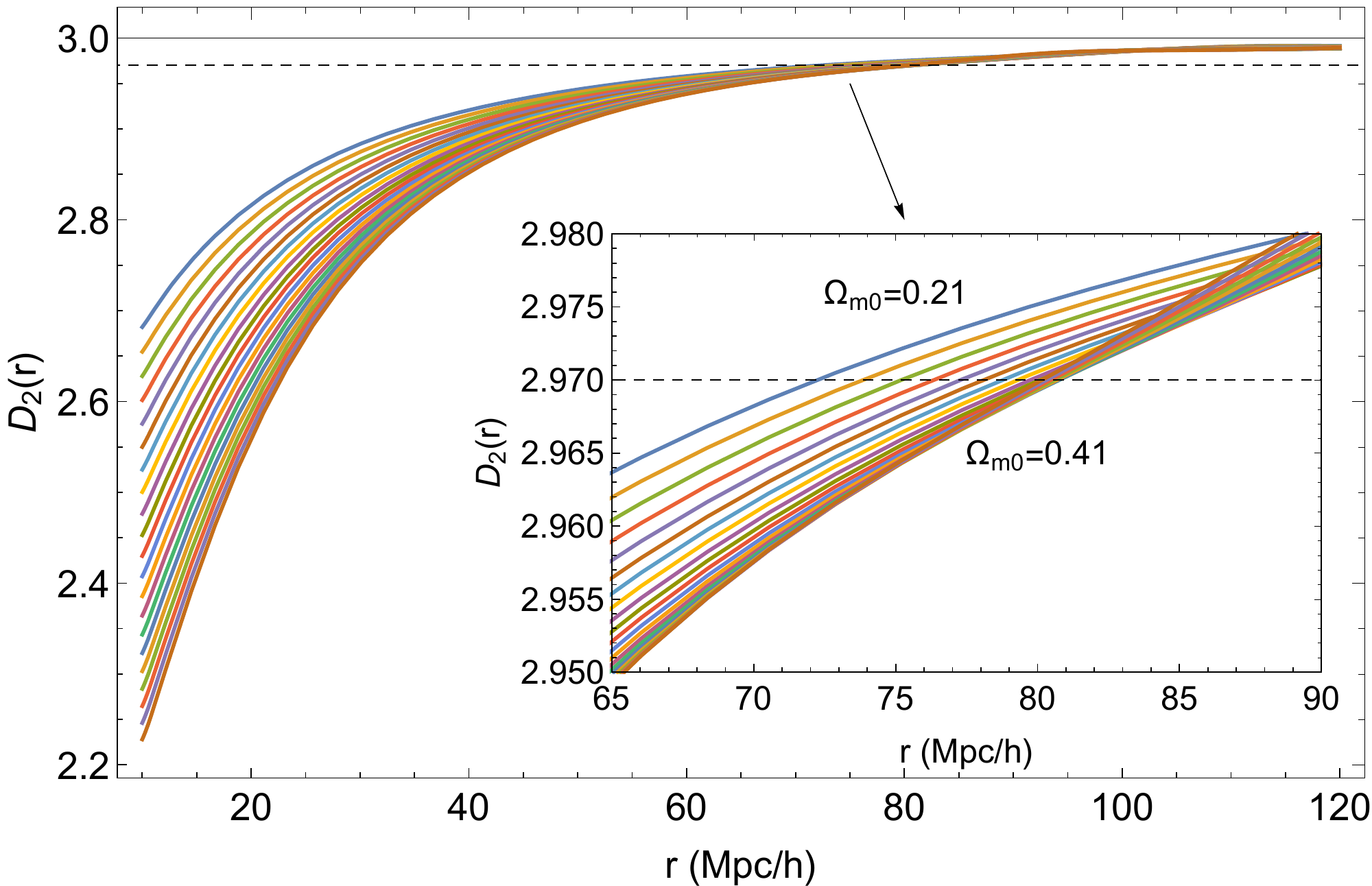}
	\includegraphics[width=0.49\textwidth]{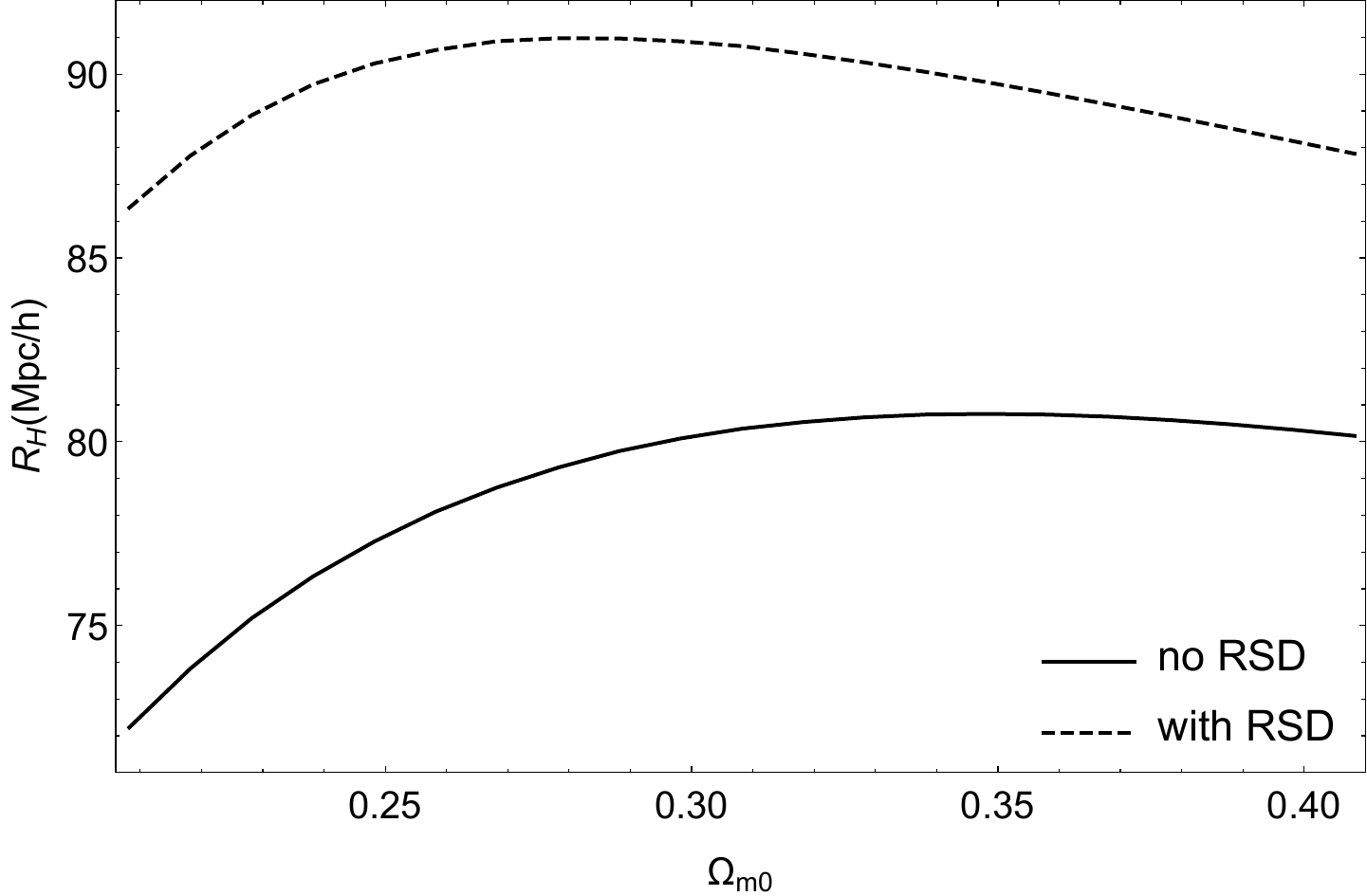}
	\caption{The fractal dimension $\mathcal{D}_2(r)$ (left) and the homogeneity scale $R_H$ (right) as a function of the present day matter density $\Omega_{m0}$. The dashed line corresponds to $1\%$ deviation from homogeneity. Clearly, the homogeneity scale $R_H$, either with RSDs or not, is not a one-to-one function for values of $\Omega_{m0}$ in that range. The cosmology used is described in the text.}
	\label{fig:hom}
\end{figure*}

The reconstruction of both quantities the BAO and the homogeneity scales can be seen in Figs.~\ref{fig:bao} and \ref{fig:hom} respectively\footnote{We note that the results shown here for the homogeneity scale remain the same if one considers the combination $R_H/D_V$.}. Specifically, in Fig.~\ref{fig:bao} we show the galaxy correlation function $\xi(r)$ (left) for $\Omega_{m0}\in[0.21,0.41]$ and the BAO scale $R_{BAO}$ (right) as a function of the present day matter density $\Omega_{m0}$.  As can be seen, the BAO scale is a monotonically decreasing function for all values of $\Omega_{m0}$ in that range. On the other hand, in Fig.~\ref{fig:hom} we show the fractal dimension $\mathcal{D}_2(r)$ (left) and the homogeneity scale $R_H$ (right) as a function of the present day matter density $\Omega_{m0}$. The dashed line corresponds to $1\%$ deviation from homogeneity. Clearly, the homogeneity scale $R_H$, is not a one-to-one function for values of $\Omega_{m0}$ in that range.

The astute reader will notice an important difference in the behavior of the homogeneity scale $R_H$ and the BAO scale $R_{BAO}$ as given in the right panels of Figs.~\ref{fig:bao} and \ref{fig:hom}. While the BAO scale $R_{BAO}$ is monotonic and shows significant change, roughly $\sim20\%$ for reasonable values of $\Omega_{m0}$, on the contrary the homogeneity scale $R_H$ is not \textit{one-to-one} and shows only small deviations, roughly $\sim4\%$ for reasonable values of $\Omega_{m0}$, thus it is rather insensitive to the value of $\Omega_{m0}$.

Moreover, we find that for the choice of the cosmology we assumed, $R_H$ has a maximum at $\Omega_{m0}\sim0.34$ thus complicating the extraction of information even further. This means that even if one was able to extract $R_H$ from a catalog (real or mock), then this value of $R_H$ would correspond to more than one value of $\Omega_{m0}$ thus inducing spurious degeneracies.

Other cosmological parameters such as $(h, w, \Omega_{\Lambda0})$ could also be considered. However, in a flat universe clearly we have $\Omega_{\Lambda0}=1-\Omega_{m0}$, so the issue of the degeneracy and not being one-to-one function remains for the dark energy density. For the Hubble parameter $h$, we expect it to just rescale the distances, while the equation of state $w(z)$ will behave similarly, albeit more weakly.

We now include RSDs and repeat our previous calculations following the analysis of Refs.~\cite{Hamilton:1997zq,Valageas:2012ex}. At large scales the main effect will be described by the Kaiser formula \cite{Kaiser:1987qv} that relates the power spectrum between real and redshift space:
\be
P^s(k)=(1+\beta \mu^2_k)^2P(k),
\ee
where $\mu_k\equiv \hat{z}\cdot \hat{k}$ is the cosine of the angle between the wavenumber $\vec{k}$ and the line of sight $\hat{z}$, while $\beta=f(a)/b$ where $f(a)=\frac{d \ln\delta}{d \ln a}$ is the linear growth and $b$ is the mean bias of the sample. Then, the power spectrum can be written in terms of Legendre polynomials as
\be
P^s(k)=\mathcal{L}_0(\mu_k) P_0^s(k)+\mathcal{L}_2(\mu_k) P_2^s(k)+\mathcal{L}_4(\mu_k) P_4^s(k),
\ee
where $\mathcal{L}_\ell(\mu_k)$ are the usual Legendre polynomials and the multipoles of the linear power spectrum are given by:
\bea
P_0^s(k)&=& \left(1+\frac23 \beta+\frac15\beta^2\right)b^2 P(k), \\
P_2^s(k)&=& \left(\frac43 \beta+\frac47 \beta^2\right)b^2 P(k),  \\
P_4^s(k)&=& \frac{8}{35} \beta^2 b^2 P(k).
\eea
The multipoles of the correlation function will then be given by \cite{Hamilton:1997zq}:
\be
\xi_\ell^s=\frac{i^\ell}{2\pi^2}\int_0^\infty j_\ell(k r) k^2 P_\ell^s(k) dk,\label{eq:corfuncmulti}
\ee
where for $\ell=0,2,4$ we have the monopole, quadrupole and hexadecapole respectively. With these in mind, we then find that the effect of the RSDs on the monopole of the correlation function is given by
\bea
\xi_0^s&=&\frac{1}{2\pi^2}\int_0^\infty j_0(k r) k^2 \left(1+\frac23 \beta+\frac15\beta^2\right)b^2 P(k) dk \nn \\
&=&\frac{1}{2\pi^2}\textrm{RSD}(\beta,b)\int_0^\infty j_0(k r) k^2 P(k) dk,\label{eq:corfuncrsd}
\eea
where $\textrm{RSD}(\beta)=\left(1+\frac23 \beta+\frac15\beta^2\right)b^2$, with $\beta=f/b$ and in order to keep the analysis simple we have assumed the bias does not depend on the scale $k$ and in any case, scale dependent bias would only affect the results in small scales. 

Using Eq.~\eqref{eq:corfuncrsd} in \eqref{eq:d2xi} we find that in the presence of RSDs the fractal dimension $\mathcal{D}_2(r)$ can be written as
\bea
\mathcal{D}_2(r)&=&3\nn\\
&+&\frac{d \ln}{d \ln r}\left[1+\frac{3}{r}\textrm{RSD}(\beta,b)\int_0^\infty \frac{1}{2\pi^2}~j_1(k r) k P(k) dk\right],\nn\\\label{eq:d2xi1rsd}
\eea
where the main difference with Eq.~\eqref{eq:d2xi1} is the presence of the $\textrm{RSD}$ term.

Using Eqs.~\eqref{eq:corfuncrsd} and \eqref{eq:d2xi1rsd} we consider  the effects of RSDs on the homogeneity scale by assuming a redshift dependent bias of the form $b=\sqrt{1+z}$. In Fig.~\ref{fig:hom} (right) we show the homogeneity scale $R_H$ without (solid black line) and with (dashed line) RSDs. As can be seen, the general behavior is still the same, except that the homogeneity scale is shifted by roughly $10~\textrm{Mpc}/h$.

\section{Comparison with the N-body simulation \label{sec:nbody}}

In this section, we will now compute the fractal dimension $\mathcal{D}_2(r)$ from a single N-Body simulation and compare with our theoretical expectations.

\begin{figure}[!t]
	\centering
	\includegraphics[width=0.49\textwidth]{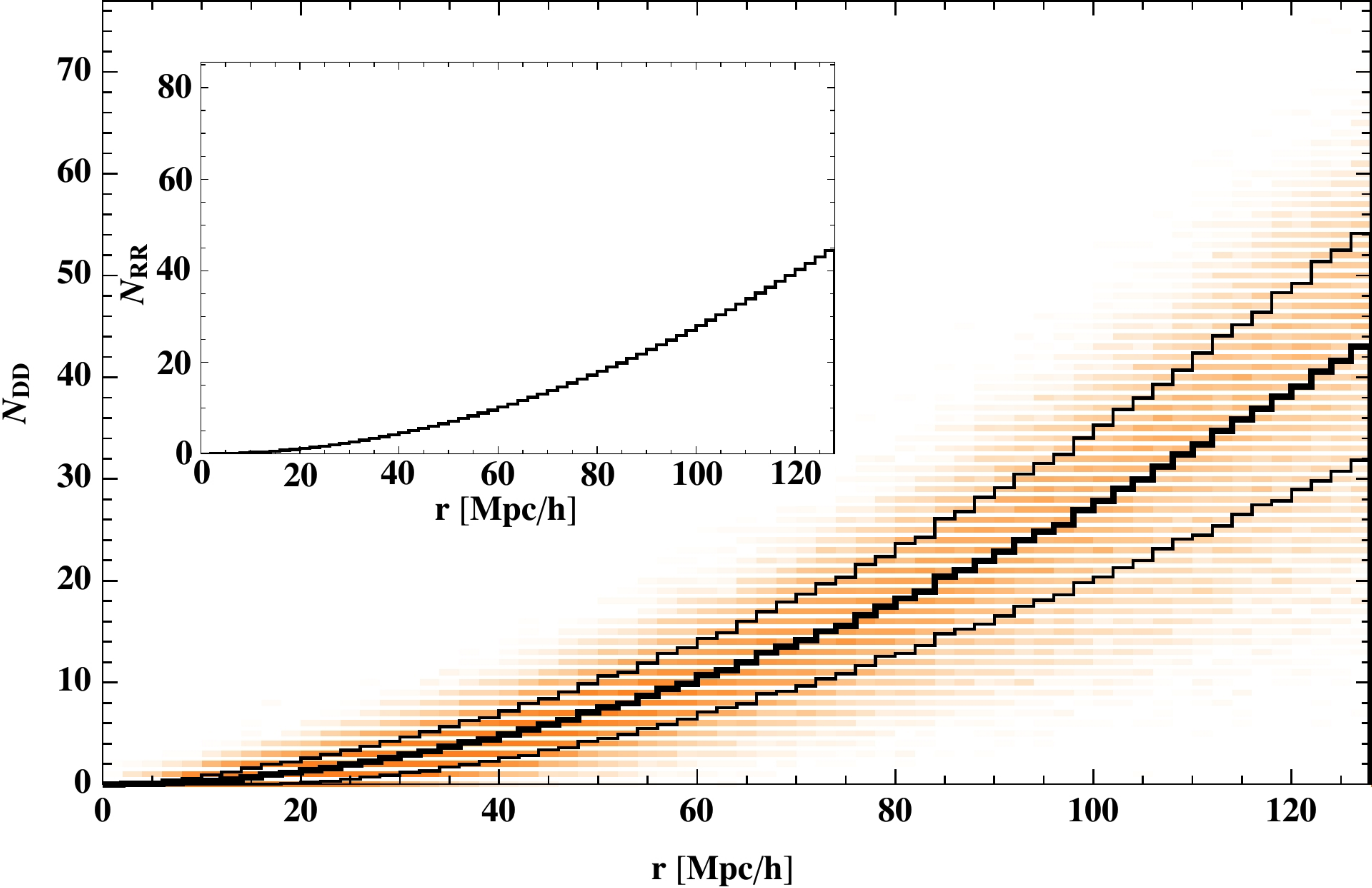}
	\caption{The particle number counts (DD) of the simulation as a function of the radius of the spherical shell up to $r_{max}=128$ Mpc/h and the expected random number counts (RR) (inset), averaged over 1024 randomly distributed centers of the shells, fully contained within the simulation box, in 64 evenly distributed bins of width $\Delta r = 2$ Mpc/h. The upper and lower black lines correspond to the $1\sigma$ confidence bands.}
	\label{fig:counts}
\end{figure}

\begin{figure*}[!t]
	\centering
	\includegraphics[width=0.49\textwidth]{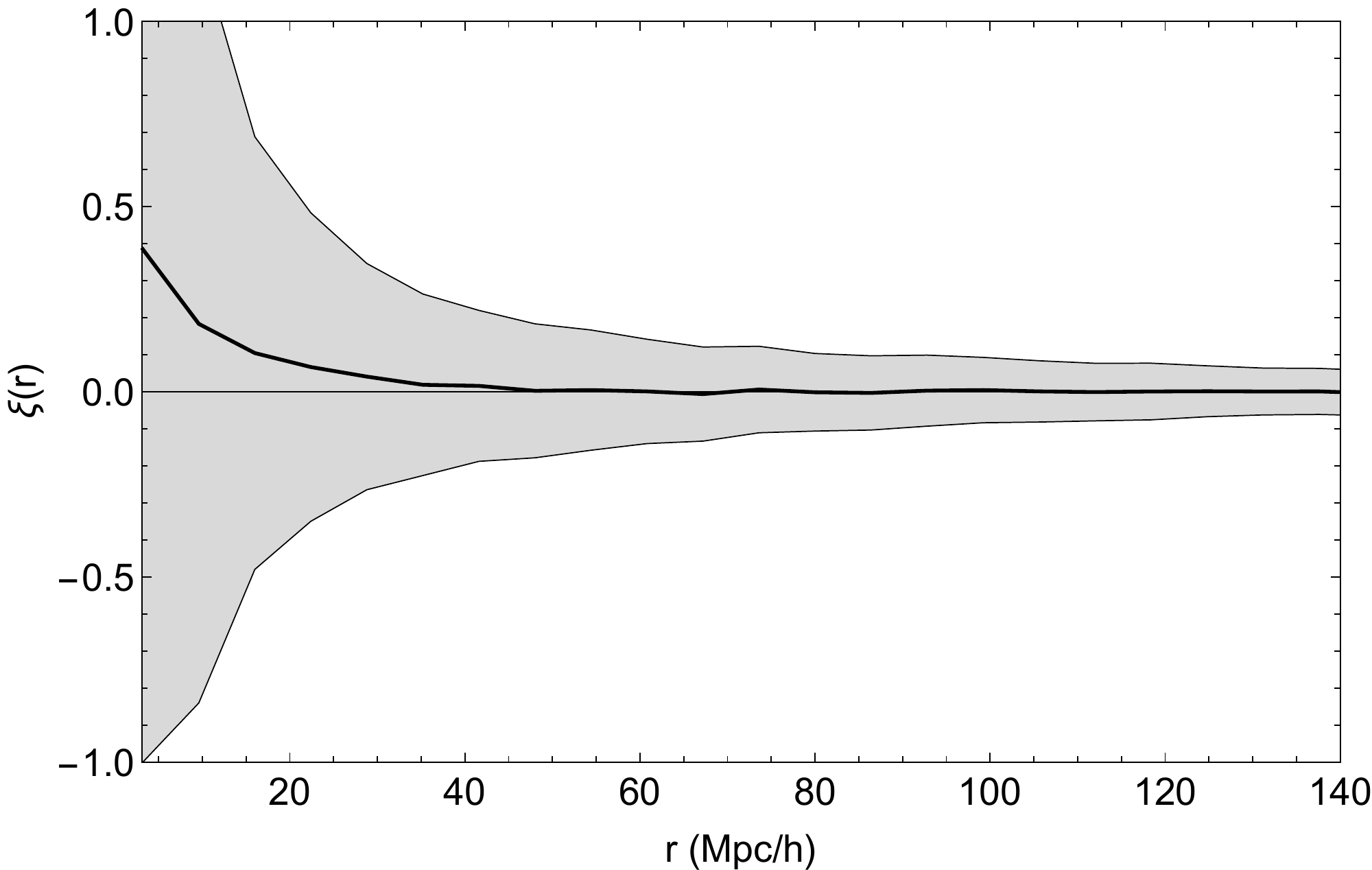}
    \includegraphics[width=0.49\textwidth]{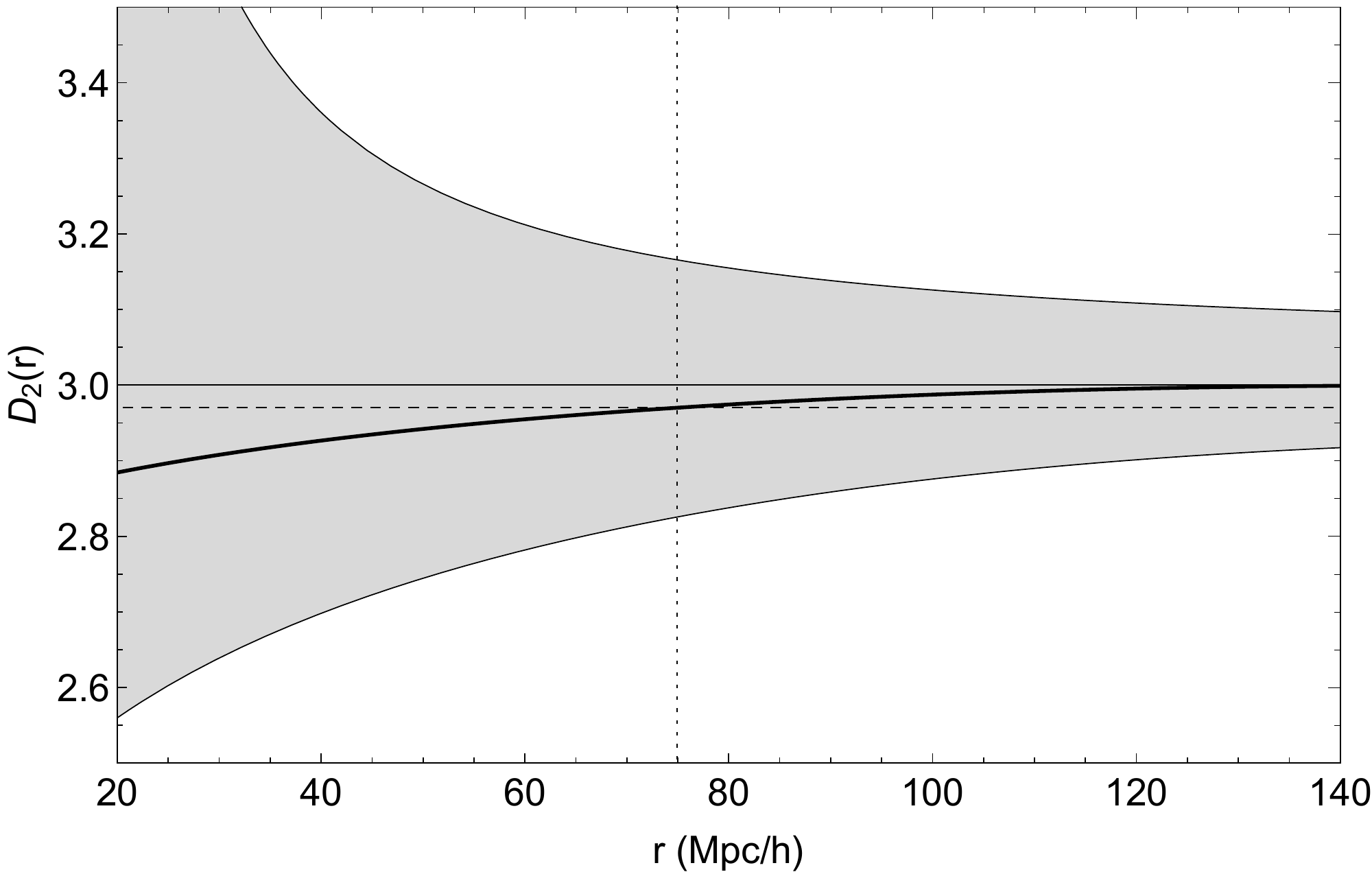}
	\caption{The galaxy correlation function $\xi(r)$ (left) and the fractal dimension $\mathcal{D}_2(r)$ (right) obtained from the N-Body simulation as described in the text. The vertical dotted line in the right panel corresponds to the homogeneity scale $R_H$, which we find to be $R_H\sim 75$Mpc/h. The horizontal dashed line indicates $1\%$ deviation from homogeneity or $\mathcal{D}_2=2.97$. In both panels, the upper and lower thin black lines correspond to the 1$\sigma$ confidence bands: for the correlation function we used linear interpolation between bins, while for the fractal dimension we used a fit of the form $\mathcal{D}_2(r)=a_0+a_1 r^n+a_2 r^{2n}$. }
	\label{fig:nbody}
\end{figure*}

Our N-Body simulation was performed with the widely used code Gadget-2 (GPL licensed) \cite{Springel:2005mi}\footnote{The Gadget-2 code can be downloaded from here \url{https://wwwmpa.mpa-garching.mpg.de/gadget/}}. The simulations start at an initial redshift of $z_{ini} = 64$ evolving until today at a redshift of $z_{end} = 0$. It is at this last redshift bin at which the fractal dimension of the dataset is calculated. The simulation has $ N_P = 2 \times 512^3$ particles divided evenly in $N_b = N_c = 512^3$, $ N_P = 2 \times 512^3 = N_b + N_c $ baryonic and cold dark matter species. The simulation box is of size $L_B = 2048$ Mpc/h and the cosmology corresponds to the flat universe of Planck 2018 best-fit \cite{Aghanim:2018eyx}: $ (\Omega_{b0},\Omega_{c0},\Omega_{\Lambda0}) = (0.0482205,0.2628795,0.6889)$. The Hubble parameter at present is set to $h\equiv \frac{H_0}{100 \textrm{km/s/Mpc}} =0.6766$.

The initial conditions (ICs) for the simulation were generated with the GPL licensed code N-GenIC \footnote{The N-GenIC code can be downloaded from here \url{https://wwwmpa.mpa-garching.mpg.de/gadget/n-genic.tar.gz}}, with a primordial power spectrum calculated with the Eisenstein and Hu parametrization of Ref.~\cite{Eisenstein:1997ik}, with parameters $n_s = 0.9665$ and normalization $\sigma_8 = 0.811$. Note that the N-body simulation does not contain RSDs, as this is an observer effect only. In order to include those, one would have to displace all particles with a factor of $\Delta \vec{x}=\frac{\vec{x}}{|\vec{x}|}\frac{1+z}{H(z)}v_r$, where $v_r$ is the radial velocity, see Ref.~\cite{Scrimgeour:2012wt} for more details.

In Fig.~\ref{fig:counts} we show the particle number counts ($N_{DD}$) histogram of the simulation as a function of the radius of the spherical shell up to a maximum distance of $r_{max}=128$ Mpc/h, as well as the expected random number counts ($N_{RR}$) histogram for all of the realizations, and their confidence bands. This we use to compute the correlation function $\xi (r)$ directly from the simulation by means of the natural estimator: $\xi (r) =  N_{DD}(r+\Delta r) / N_{RR}(r+\Delta r) - 1$, where $\Delta r$ is the bin width.

Finally, in Fig.~\ref{fig:nbody} we show the galaxy correlation function $\xi(r)$ (left) and the fractal dimension $\mathcal{D}_2(r)$ (right) obtained from our N-Body simulation. The vertical dotted line in the right panel corresponds to the homogeneity scale $R_H$, while the horizontal dashed line indicates $1\%$ deviation from homogeneity or $\mathcal{D}_2=2.97$. Clearly, the result is in good agreement with the theoretical predictions from the previous section.

The gray-shaded areas in Fig.~\ref{fig:nbody} correspond to the $1\sigma$ errors that were obtained by considering all different spheres of size $R=200 \textrm{Mpc}/h$ inside our simulation box, which is of size $L_B = 2048$ Mpc/h and contains approximately $\left(\frac{L_B}{2R}\right)^3\sim 130$ spheres in a regular lattice packing. Our errors are much larger than those of Ref.~\cite{Scrimgeour:2012wt} mainly due to the smaller number of sphere centers within the simulation box we considered and also the rather low resolution, and the use of the natural estimator which has by default larger errors, both due to computational constraints.

\section{Conclusions \label{sec:conclusions}}
In this paper we performed a comparison of the BAO and homogeneity scales in order to assess the suitability of the latter as a standard ruler, as was proposed in Ref.~\cite{Ntelis:2018ctq}.

We found that while the BAO scale is monotonic and depends strongly on the matter density parameter $\Omega_{m0}$, changing roughly by $\sim20\%$ for reasonable values of $\Omega_{m0}$, the homogeneity scale not only is not \textit{one-to-one}, but it is also rather insensitive to the value of $\Omega_{m0}$, changing only by $\sim4\%$ in that range of values. Furthermore, the homogeneity scale $R_H$ has a maximum at $\Omega_{m0}\sim0.34$, which means that even if one was able to extract $R_H$ from a catalog (real or mock), then this value of $R_H$ would correspond to more than one value of $\Omega_{m0}$ thus inducing spurious degeneracies.

Moreover, we found similar results when including the redshift space distortions. Specifically, the general behavior for the homogeneity scale is still the same, except that now it is shifted by roughly $10~\textrm{Mpc}/h$ and the maximum in terms of the matter density happens at $\Omega_{m0}\sim0.28$. Clearly, in either case, with or without RSDs, the main conclusions of our analysis are the same.

Finally, in order to validate our theoretical predictions, we also compared our results with those obtained from an N-body simulation. We found that the prediction for the homogeneity scale from the N-body simulation was $R_H\sim 75$Mpc/h, which is compatible with the theoretical predictions from the previous section, as seen in Fig.~\ref{fig:nbody}.

To summarize, we find that the homogeneity scale $R_H$ does not have the desired properties to be a useful cosmological probe: being an one-to-one function with respect to the cosmological parameters of interest and having a strong dependence on cosmology, both attributes similar to those of the BAO scale. Given that Ref.~\cite{Ntelis:2018ctq} assumed the general case of a non-flat Universe, then our results are complementary to their finding. However, in our more restricted cosmology of a flat $\Lambda$CDM Planck 2018 best-fit cosmology, we conclude that $R_H$ cannot be considered a suitable standard ruler.

\section*{Acknowledgements}
The authors would like to thank J.~Garc\'{\i}a-Bellido, J.M.~Le Goff, R.~James and P.~Ntelis for useful discussions. The authors also acknowledge support from the Research Project No. FPA2015-68048-03-3P [MINECO-FEDER], the Centro de Excelencia Severo Ochoa Program SEV-2016-0597 and use of the Hydra cluster at the IFT. S.N. acknowledges support from the Ram\'{o}n y Cajal program through Grant No. RYC-2014-15843. M.T. acknowledges support from the Grant No. BES-2016-077817 [MINECO-FPI] .

\bibliography{homogeneity}

\end{document}